\documentclass[aip,cha,reprint,nofootinbib,a4paper]{revtex4-1}

\setcitestyle{numbers,square}
\usepackage{color}
\usepackage[]{graphicx}
\usepackage{latexsym}
\usepackage{natbib}
\usepackage{amsmath}
\usepackage[caption=false]{subfig}
\usepackage{multirow}
\usepackage{mathtools}
\usepackage{textcomp}
\usepackage{booktabs}
\usepackage{comment}

\usepackage[pdftex,hyperfigures,breaklinks,colorlinks,citecolor=black,linkcolor=black]{hyperref} 

\newcommand{\be}{\begin{equation}}
\newcommand{\ee}{\end{equation}}
\newcommand{\bc}{\begin{cases}}
\newcommand{\ec}{\end{cases}}

\begin{document}

\title{What shifts threshold distributions in social contagions?}

\author{Luca Lazzaro$^*$}
\affiliation{University of Zurich}
\affiliation{University Research Priority Program on Social Networks}

\author{Manuel S. Mariani}
\affiliation{University of Zurich}
\affiliation{University Research Priority Program on Social Networks}
\affiliation{Corvinus University of Budapest}

\author{René Algesheimer}
\affiliation{University of Zurich}
\affiliation{University Research Priority Program on Social Networks}

\author{Radu Tanase$^*$}
\affiliation{University of Zurich}
\affiliation{University Research Priority Program on Social Networks}

\begin{abstract}
Individual thresholds in social contagions capture what fraction of others must adopt a new product or behavior before an individual adopts it. Yet in a given choice context and population, we rarely know the empirical threshold distribution or what explains its heterogeneity. This limits the ability to predict diffusion dynamics and design effective network interventions. We address this by measuring thresholds with an incentivized online experiment in which participants make adoption decisions for products that vary in attractiveness and payoff uncertainty. The results show that product characteristics shift threshold distributions: lower attractiveness and higher uncertainty increase thresholds. Individual characteristics, however, remain essential for explaining threshold heterogeneity. We then use data-driven simulations to show that both product characteristics and individual characteristics moderate the effect of long ties in social contagions. In sum, the threshold distribution is the behavioral microfoundation of social contagion models and identifying empirical correlates of thresholds is crucial for understanding how network structure affects social contagions.
\end{abstract}

\maketitle

\onecolumngrid
\vspace{-2.5em}
\begin{center}
    \parbox{0.85\textwidth}{%
        \footnotesize{\textsuperscript{*}To whom correspondence should be addressed.
        Email: \url{luca.lazzaro@uzh.ch} or \url{radu.tanase@uzh.ch} 
        }
    }
\end{center}
\vspace{1em}

\twocolumngrid
Choices are often interdependent. Whether adopting a new practice, following a social norm, or purchasing a product, individuals can be influenced by the choices of others, giving rise to social contagion \cite{coleman1957physicians, christakis2007thespread, banerjee2013thediffusion, aral2017excercisecontagion}. This interdependence can be captured by an individual's threshold: the fraction of adopting peers required before they adopt a behavior themselves \cite{granovetter1978threshold}. Aggregated across a population, these individual thresholds form a threshold distribution, which provides the behavioral microfoundation of social contagion models \cite{centola2007complex, eckles2024long, wan-2025diffusion}. Therefore, the threshold distribution is the natural bridge between behavioral research on decision-making and theoretical research on social contagion. Yet we rarely observe this distribution empirically, and we know even less about what causes it to shift across settings. This gap limits our ability to predict diffusion dynamics and design effective network interventions \cite{valente2012network}.

Decision-makers can accelerate or slow diffusion by changing who is connected to whom. Among the most studied of these interventions is the introduction of long ties---connections between individuals with few or no mutual contacts \cite{granovetter1973strength}. Long ties can either accelerate \cite{granovetter1973strength, brown1987social, burt1992structural, watts1998collective, goldenberg2001talk, lu2011small, bakshy2012role, lee2020strength, rajkumar2022causal, jahani2023long} or impede \cite{centola2007complex, centola2010spread, zheng2013spreading, guilbeault2018complex, notarmuzi2022universality, mariani2024collective} diffusion. In social contagion models, this distinction can be traced to the threshold distribution. When a single adopting peer is sufficient to trigger adoption, long ties accelerate diffusion by reaching otherwise unexposed parts of the network, a regime known as simple contagion \cite{granovetter1973strength, dodds2005generalizedmodel}. In contrast, when adoption requires reinforcement from multiple peers, long ties reduce the local clustering needed for reinforcement, causing diffusion to stall. This regime is known as complex contagion \cite{centola2007complex, centola2010spread, guilbeault2018complex}. Recent theoretical work shows that even small departures from the classic complex contagion model, such as introducing stochasticity into adoption decisions, are sufficient to restore the advantage of long ties \cite{eckles2024long, sassine2024does, wan-2025diffusion}. These theoretical results demonstrate the central role of the threshold distribution and show that small changes can reverse the effect of long ties. However, they all rely on assumed threshold distributions rather than empirically estimated ones.

Only a few studies have measured threshold distributions \citep{ludemann1999thresholds, tanase2024integrating, janas2026eliciting}. Those studies find that threshold distributions are heterogeneous. Some individuals adopt after exposure to a single peer; others adopt only under reinforcement from many.  Simple and complex adoption are therefore not competing accounts but two regions of a single threshold distribution \cite{andres2025distinguishing}. What previous work does not address, and what theoretical models are not designed to answer, is why one setting yields a lower-threshold distribution and another a higher-threshold distribution. We move beyond measuring thresholds to identifying what shifts them.

We conduct an incentivized online experiment in which participants make adoption decisions for products while observing varying numbers of adopting peers. We elicit each participant's adoption threshold. We manipulate two product characteristics: attractiveness and payoff uncertainty. The two characteristics operate through distinct channels. Product attractiveness changes the expected payoff of adoption. Payoff uncertainty changes the informational value of others' choices. The setting captures common features of many adoption decisions: adoption yields an uncertain payoff, while others’ choices may provide informative signals \cite{feldmanhall2019resolving,Turner2023uncertaintyform, chen2025sequential-learning}.

In this paper, we identify what shapes the threshold distribution and show that these same characteristics predict the magnitude of the effect of long ties in social contagions. Lower product attractiveness and higher payoff uncertainty shift the threshold distribution toward higher thresholds, where individuals require more adopting peers before adopting. Individual characteristics, partly captured by risk aversion and pessimistic beliefs about a product's success, predict the distribution dispersion. Embedding these measured distributions in network simulations, we find the magnitude of the effect of long ties is moderated by the same product and individual characteristics. What is framed as a debate between simple and complex contagion is at root a question about the threshold distribution, a behavioral quantity to be measured and explained.

\begin{figure}
\centering
\includegraphics[width=1\linewidth]{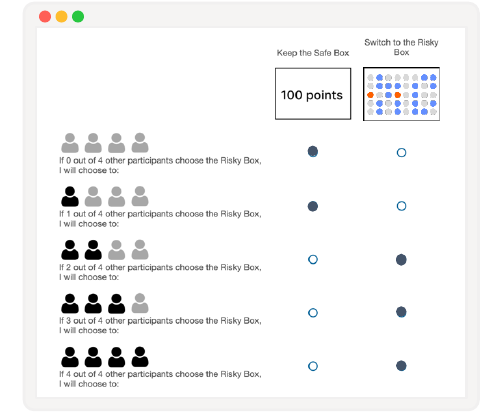}
\caption{\textbf{Eliciting adoption thresholds with the strategy method.} In the experiment, participants decide whether to adopt products varying in attractiveness and payoff uncertainty. Participants provide a separate adoption decision for each possible peer adoption level (0–4 adopting peers out of 4, presented in an increasing order). Every adoption decision is incentive aligned (see Methods). This procedure elicits each participant's adoption threshold---the minimum fraction of adopting peers at which the individual switches from non-adoption to adoption. The example in the figure indicates a threshold of one out of four ($1/4$).}
\label{fig:framework}
\end{figure}

\begin{figure*}
\centering
\includegraphics[width=0.9\linewidth]{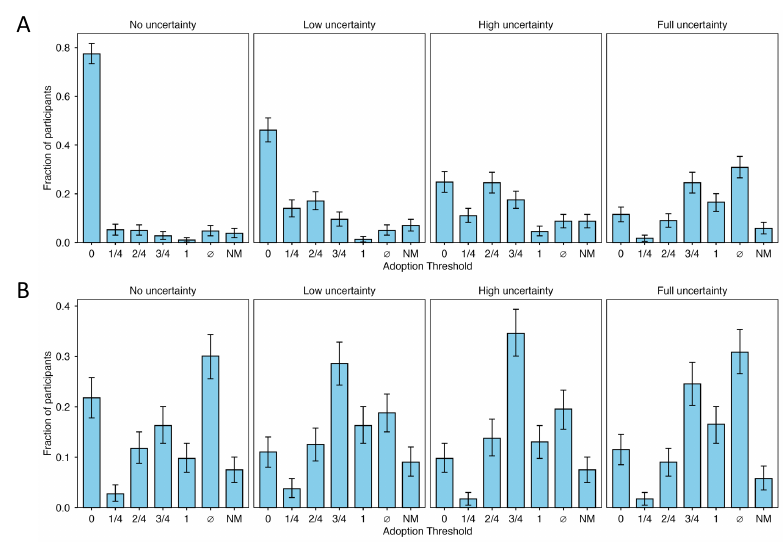}
\caption{\textbf{Empirical threshold distributions across attractiveness and uncertainty conditions.}  (A) High attractiveness condition; (B) low attractiveness condition. Within each panel, bars show the proportion of participants ($n = 399$) falling into each threshold category for each uncertainty level. Error bars  denote $95\%$ bootstrap confidence intervals, computed by resampling participants with replacement ($2,000$ resamples). A threshold is the minimum fraction of adopting peers at which an individual switches from non-adoption to adoption, where $m\in\{0,\dots,4\}$ is the number of adopting peers out of $k=4$, derived from each participant's experimental choices: $0=$ unconditional adopter (adopts regardless of peer behavior); $1/4$, $2/4$, $3/4$, $1=$ conditional adopters requiring progressively more peers; $\emptyset=$ unconditional non-adopter (never adopts, i.e.\ does not adopt at any $m$); \texttt{NM}$=$ non-monotonic adoption pattern (not representable by a single threshold). Under full uncertainty, the attractiveness manipulation is inoperative by design; the  corresponding distributions are identical across panels.}
\label{fig:threshold-distributions}
\end{figure*}

\subsection*{Eliciting individual adoption threshold}
Measuring thresholds is challenging. A threshold is a contingent quantity, the fraction of adopting peers at which an individual switches to adopting. In real-world diffusions, each individual is observed at a single realized level of peer adoption, so the threshold cannot be identified from direct response \cite{manski1993identification, berry2018opacity}. One can recover an average threshold across a population \cite{leskovec-2007viral, bakshy2012social, lee2022complex}, but averaging discards the shape of the distribution, and this hidden heterogeneity would make simulations of collective dynamics unreliable \cite{st2024nonlinear}.

Individual measurement avoids this identification problem. Existing approaches recover individual thresholds from a reported switch point \cite{janas2026eliciting}, from responses to hypothetical scenarios in choice experiments \cite{tanase2024integrating}, or from observational data under a parametric model \cite{tran2022heterogeneous}. All the existing approaches impose a threshold rule on individual choices. We use the strategy method \cite{selten1967strategiemethode} to elicit the full contingent choice profile, that is, an adoption decision at every peer adoption count. Previous work found that contingent and direct-response elicitation produce similar behavior across domains \cite{brandts2011strategy, jordan2015effect}, supporting the use of the strategy method. In the experiment, participants indicate, for each possible number of adopting peers (0--4), whether they would adopt (Fig.~\ref{fig:framework}). This procedure recovers one contingent choice profile per participant--product pair. For monotonic profiles, the smallest fraction of adopting peers at which the participant switches to adoption defines the individual threshold. Non-monotonic profiles have no single threshold and are treated as general adoption rules.

Our experimental setting is one of informational social learning. Participants make adoption decisions conditional on peers' (other participants) decisions, and the fact that different participants may receive different information about the product is common knowledge. This setting makes peer adoption decisions potentially informative signals of product quality \cite{banerjee1992simple, bikhchandani1992atheory, chen2025sequential-learning}.

Studying what shifts the threshold distribution requires a setting in which product characteristics can be varied while keeping adoption decisions consequential. We model products as urn lotteries in the tradition of Ellsberg \cite{ellsberg1961risk}, a paradigm with two properties essential to our design. First, it permits independent manipulation of the two product characteristics we study: attractiveness and payoff uncertainty. Second, it supports incentivized elicitation. Participants' adoption decisions have real monetary consequences, reducing concerns about purely hypothetical responses. Each product is visualized as an urn containing 40 balls (Fig.~\ref{fig:lottery}). Adopting the product corresponds to drawing a ball at random from the urn, yielding 300 points if the ball is blue and 0 points if it is orange. Choosing not to adopt provides a guaranteed payoff of 100 points.

We manipulate two product characteristics: product attractiveness and payoff uncertainty. In this setting, product attractiveness refers to the visible payoff advantage of adoption relative to non-adoption. We operationalize attractiveness through the ratio of blue to orange balls in the visible portion of the urn. We create two attractiveness levels: 90\% of blue balls (high attractiveness) and 50\% of blue balls (low attractiveness). Holding beliefs about the hidden balls constant, the high-attractiveness product has a higher expected payoff than the low-attractiveness product. Payoff uncertainty captures how much of the product's winning probability is unknown. We operationalize uncertainty by hiding the colors of a fraction of the balls (displaying them as gray) at four levels: 0\% (no uncertainty), 25\% (low uncertainty), 50\% (high uncertainty), and 100\% (full uncertainty). We expect these manipulations to create a wide range of threshold distributions encompassing both simple and complex contagion processes \cite{guilbeault2018complex}.


The design yields seven unique product configurations because under full uncertainty, all balls are hidden, so the visible information contains no basis for distinguishing high- from low-attractiveness products (see Methods and Fig.~\ref{fig:lottery} for full details). Participants ($n=399$) complete all seven product configurations in randomized order, with decisions incentivized (see Methods). We additionally measure risk preferences using the multiple price list method \cite{holt2002risk} and elicit subjective probability of success for uncertain products (see Methods).

\section*{Results}
We organize the results around the paper's central question: what shifts threshold distributions? The analyses proceed in three steps. First, we show that experimentally elicited adoption thresholds vary systematically with product characteristics, while retaining substantial heterogeneity within every condition. Second, we examine whether this residual heterogeneity is associated with participant-level characteristics, including beliefs and preferences. Third, we embed the empirically measured threshold distributions in network simulations to examine the effect of long ties in social contagions.

\begin{figure*}
\centering
\includegraphics[width=0.7\linewidth]{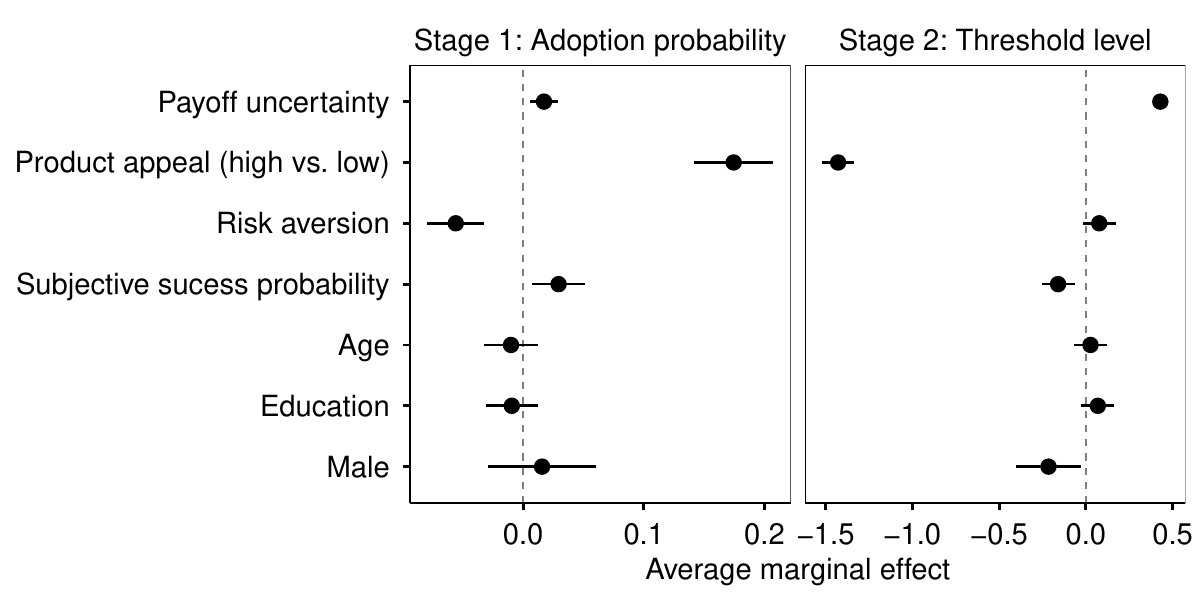}
\caption{\textbf{Correlates of adoption thresholds.} Average marginal effects from a two-stage mixed-effects model. Both stages include participant random intercepts to account for repeated measures across products. Stage~1 (left): binary logistic regression estimating the change in adoption probability ($N_{\text{obs}} = 1{,}956$ participant--product observations from $N_{\text{participants}} = 326$). Stage~2 (right): linear regression estimating the change in adoption threshold, conditional on adoption ($N_{\text{obs}} = 1{,}670$ from $N_{\text{participants}} = 320$; six participants who never adopted across any product contribute to Stage~1 only). The sample includes participants with monotonic threshold responses across all products. The full-uncertainty product is excluded from both stages because product attractiveness was not experimentally varied in that condition. Product attractiveness is a binary factor (higher vs.\ lower); all other continuous predictors are standardized (one-SD units). Unstandardized effects for experimental manipulations are reported in the main text. Points indicate point estimates; horizontal lines indicate 95\% confidence intervals. For Stage~2, fixed effects explain 34.9\% of threshold variance (marginal $R^2$); including participant random intercepts raises this to 60.3\% (conditional $R^2$).}
\label{fig:reg}
\end{figure*}

\subsection*{Product characteristics shift the threshold distribution}
Most response patterns are monotonic (92.9\%), meaning that adoption decisions are non-decreasing in the number of adopting peers and can therefore be described by a threshold. Fig.~\ref{fig:threshold-distributions} shows that the elicited threshold distributions are wide and their shape shifts with product characteristics. Across all seven products, thresholds cover the full range of possible values. Importantly, every distribution retains a fraction of low-threshold individuals. We describe how product characteristics shift the threshold distributions using three measures.

First, we compute the average threshold among adopters (measured in peer-count units 0--4). The average threshold captures the average number of peers required before adopting the product. Lower attractiveness increases the average adoption threshold. For example, under no uncertainty, the average threshold increases from $0.30$ for high-attractiveness products to $1.83$ for low-attractiveness products. Higher uncertainty also increases average thresholds. For high-attractiveness products, the average threshold rises from $0.30$ under no uncertainty to $1.59$ under high uncertainty.

Second, we compute the fraction of low-threshold individuals (threshold of 0/4 or 1/4). The fraction of low-threshold individuals is relevant for the role of long ties in social contagions \cite{eckles2024long}. This fraction decreases under higher uncertainty and lower attractiveness. For example, among high-attractiveness products, the fraction of low-threshold individuals falls from 82.7\% under no uncertainty to 35.8\% under high uncertainty. Nonetheless, the fraction of low-threshold individuals is non-zero for all products.

Third, we compute the fraction of conditional adopters, individuals who change their adoption decision depending on the choice of others (threshold of 1/4, 2/4, 3/4 or 4/4). In contrast, unconditional (non-) adopters describe individuals who (never) always adopt the product regardless of peer choices. Lower attractiveness and higher uncertainty generally produce a higher fraction of conditional adopters. For example, among high-attractiveness products, 14.0\% of individuals are conditional adopters under no uncertainty, compared with 57.6\% under high uncertainty. This share then falls to 51.9\% under full uncertainty as the distribution shifts toward unconditional non-adoption. For detailed results of the three measures, see Supplementary Information, Fig.~\ref{fig:t-summary}.

Together, these patterns show that product characteristics shift the threshold distribution. Less attractive and more uncertain products generate threshold distributions with higher thresholds. However, every product retains substantial heterogeneity, including a persistent fraction of low-threshold individuals. We next ask whether this residual heterogeneity is associated with differences across participants.

\begin{figure*}
\centering
\includegraphics[width=0.9\linewidth]{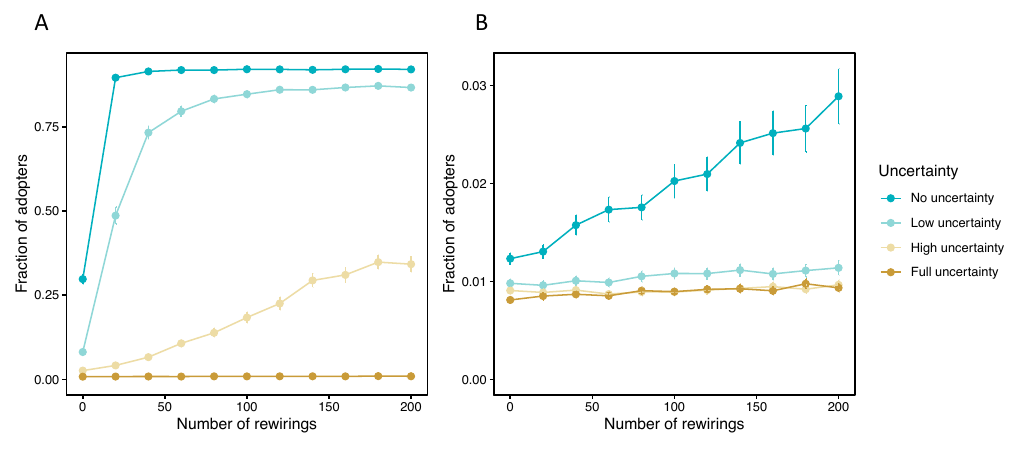}
\caption{\textbf{The effect of long ties on diffusion.} Average final fraction of
adopters as a function of the number of rewired edge pairs, for high-attractiveness
(A) and low-attractiveness (B) products; color denotes the payoff-uncertainty
condition (no $=0\%$ hidden balls, low $=25\%$, high $=50\%$, full $=100\%$).
Simulations use ring lattices ($N=399$, degree $k=4$) with degree-preserving
rewiring \cite{maslov2002specificity}; diffusion is seeded from a randomly chosen
connected pair, and thresholds are resampled with replacement and randomly assigned
to nodes each run. Points are means across $R=500$ realizations; error bars are
$95\%$ confidence intervals for the mean and are often smaller than the markers.}
\label{fig:macro_results}
\end{figure*}

\subsection*{Individual characteristics predict threshold heterogeneity}

To examine what explains heterogeneity in adoption thresholds, we estimate a two-stage mixed-effects model with participant random intercepts. The first stage estimates adoption probability using a logistic regression. The second stage estimates adoption thresholds (in peer-count units, 0--4) conditional on adoption using a linear mixed-effects model. Figure~\ref{fig:reg} displays average marginal effects; the complete model is reported in Supplementary Information, Table~\ref{tab:hurdle-main}. Continuous predictors are standardized in the figure for visual comparability; experimental manipulations are reported in natural units in the text.

Product characteristics have the largest effects on both adoption probability and threshold. High attractiveness raises adoption probability by $17.4$ percentage points relative to low attractiveness ($95$\% CI [$14.2$, $20.7$]) and, among adopters, reduces the number of peers required before adoption by $1.43$ out of $4$ ($95$\% CI [$1.34$, $1.52$]). Moving from no uncertainty to high uncertainty increases adoption probability by $4.2$ percentage points ($95$\% CI [$1.4$, $7.1$]) and, conditional on adoption, increases the required number of adopting peers by $1.05$ out of $4$ ($95$\% CI [$0.94$, $1.16$]). Uncertainty therefore has a dual effect on adoption: it raises the probability that participants adopt and their adoption threshold.

Within-product thresholds vary widely. Fixed effects explain $34.9$\% of threshold variance (marginal $R^2$). Including participant random intercepts raises the explained variance to $60.3$\% (conditional $R^2$). This $25.4$ percentage-point increase indicates substantial participant-level heterogeneity in thresholds.

Measured individual characteristics explain part of these participant-level differences. Among adopters, a one-standard-deviation increase in the subjective probability of success lowers the threshold by $0.16$ peers out of $4$ ($95$\% CI [$0.06$, $0.26$]). Participants identifying as male require $0.22$ fewer adopting peers out of $4$ than other participants ($95$\% CI [$0.03$, $0.41$]). Risk aversion reduces adoption probability by $5.6$ percentage points per standard deviation ($95$\% CI [$3.2$, $7.9$]). Age and education show small effects on thresholds, with confidence intervals spanning zero. Individual characteristics therefore help predict where participants fall within the threshold distribution, but they do not fully account for the remaining heterogeneity.

Taken together, these results show that threshold distributions are shaped by both product and individual characteristics. Product characteristics shift the location of the distribution, while participant-level heterogeneity drives the dispersion within each product condition.

\subsection*{Implications for the effect of long ties}

We next ask what the empirically measured threshold distributions imply for the effect of long ties when these distributions are used as input to a standard social contagion simulation. The simulation does not test which contagion model is correct; it propagates the elicited distributions forward and quantifies how the effect of long ties varies across the empirical regime our manipulations generate. To test this, we embed participants' experimentally elicited thresholds as decision rules in an agent-based model of social contagion. Starting from a ring lattice ($N = 399$, $k = 4$), we progressively introduce long ties by rewiring pairs of edges while preserving node degrees \cite{maslov2002specificity}. We conduct $500$ diffusion runs for each product and rewiring level. In each run, we resample participants' elicited thresholds with replacement to account for sampling variability in the threshold distribution. This procedure also incorporates variability due to network randomization and initial seeding; full details are reported in the Methods.

Figure~\ref{fig:macro_results} shows that the effect of long ties depends on product characteristics. For the high-attractiveness products (Fig.~\ref{fig:macro_results}A) under no, low, and high uncertainty, adding long ties monotonically increases the final fraction of adopters. These patterns are consistent with the interpretation that low-threshold individuals allow distant exposures to initiate diffusion in otherwise unexposed regions of the network. In these conditions, long ties benefit the diffusion even though they dilute local reinforcement for higher-threshold individuals.

The low-attractiveness products (Fig.~\ref{fig:macro_results}B) clarify the limiting case of this mechanism. Although these products still contain some low-threshold individuals, their threshold distributions are shifted far enough toward higher values (Fig.~\ref{fig:threshold-distributions}) that diffusion remains negligible across rewiring levels. In this case, the structural advantage of long ties cannot be realized, because too few individuals adopt for diffusion to extend beyond the seed. The same logic applies under full uncertainty, where diffusion stalls shortly after seeding and long ties confer no advantage.

Similar results appear not only across products, but also across populations. Holding the product fixed, we repeat the simulations after resampling thresholds from participants with different levels of risk aversion. The effect of long ties varies across these populations. Populations composed of less risk-averse participants contain more low-threshold individuals and exhibit larger gains from long ties. Populations composed of more risk-averse participants have higher thresholds and show weaker gains from long ties (Supplementary Information, Fig.~\ref{fig:pop-comp}). This population-level analysis shows that the effect of long ties depends not only on the product, but also on the composition of the population.

Across our empirical threshold distributions, long ties do not hinder diffusion. However, the magnitude of the effect depends on where product and population characteristics place the threshold distribution. The benefit of long ties is largest at intermediate distributions, where low-threshold individuals seed cascades and long ties carry adoption into unexposed parts of the network. The effect of long ties shrinks toward zero at both extremes: where thresholds are uniformly high and diffusion stalls, and where thresholds are uniformly low and the product diffuses fully (Supplementary Information, Fig.~\ref{fig:magnitude}). The results are robust to network degree, network size, and seed size (Supplementary Information, Figs.~\ref{fig:rob-degree}, \ref{fig:rob-netsize} and \ref{fig:rob-seed}).

Ring lattices provide an established framework for testing the effect of long ties. Real social networks, however, combine long ties with degree heterogeneity and unequal clustering. We repeat our main simulations on empirical networks to test whether our results hold on realistic network structures. Our results replicate; product characteristics moderate the effect of long ties on empirical networks. Yet, because empirical networks already contain long ties, adding long ties has little effect for no- and low-uncertainty products. These products already diffuse close to the ceiling in the original empirical networks, leaving little room for further improvement through rewiring. Thus, in Fig.~\ref{fig:macro_results}A, empirical networks are effectively located at a rewiring level where diffusion has already reached the ceiling. In contrast, rewiring yields the largest gains for the high-uncertainty product, where final adoption increases with further rewiring (see Supplementary Information, Sections~\ref{si:sim-empirical-selection}, \ref{si:rob-empirical} and Fig.~\ref{fig:rob-empirical}).

\section*{Discussion}
Threshold distributions are the behavioral microfoundation of social contagion models. Yet studying them remains challenging. We identify product (attractiveness, payoff uncertainty) characteristics that shift the threshold distribution and individual characteristics (risk preferences, beliefs) that predict the distribution’s dispersion. We show that both individual and product characteristics moderate the effect of long ties in social contagions.

Product attractiveness and payoff uncertainty shift the location of the threshold distribution. When more of the payoff probability is hidden, thresholds rise, consistent with evidence that uncertainty increases reliance on social information \cite{toyokawa2019social, zimmermann2023adoption}. This identifies one condition under which the high-threshold distributions assumed by complex contagion models can arise \cite{centola2007complex, guilbeault2018complex}. Attractiveness shifts the distribution even when the winning probability is known. In this case, peer choices cannot convey any new information about the probability of winning. Nevertheless, participants continue to rely on peer choices. Several mechanisms would produce this pattern: the decision may sit near indifference, so that any signal breaks the tie \cite{tversky1992choice}; participants may be uncertain about their own valuation of the product and read others' choices as evidence about it \cite{festinger1954theory}; or they may conform to the majority \cite{asch1956studies}.

Individual differences predict the dispersion of the distribution. Holding product characteristics constant, thresholds vary across individuals: some individuals require more adopting peers than others across products \cite{bryan2021hetero}. Risk preferences and subjective probability of success partly explain this heterogeneity. Risk-averse and more pessimistic individuals require more peers before adopting. Much of this heterogeneity remains unexplained, consistent with recent work documenting systematic individual differences in how individuals integrate social information \cite{molleman2020strategies} and respond to uncertainty \cite{zimmermann2023adoption}. 

Our experimental design produces threshold distributions that are heterogeneous with substantial mass among both unconditional adopters and unconditional non-adopters. Yet, theoretical work often represents threshold heterogeneity using smooth parametric forms such as normal or truncated-normal distributions \cite{centola2007complex, wan-2025diffusion}. Our results suggest that studying individual heterogeneity in threshold distribution is an important next step for understanding collective dynamics \cite{tanase2024integrating, janas2026eliciting}.

Embedding these empirical distributions in network simulations shows that the effect of long ties varies with the product and individual characteristics. When low attractiveness or high uncertainty shifts the distribution toward high thresholds and unconditional non-adoption, the benefit of long ties narrows. In the same way, simulated populations composed of more risk-averse individuals show a reduced benefit from introducing long ties. Crucially, this depends on the initial network structure. For empirical networks, which already include long ties, the largest gains are at intermediate to high uncertainty levels. Our results are consistent with the reach-reinforcement trade-off of long ties occurring within a single diffusion process \cite{wan-2025diffusion}. Given that the threshold distributions are heterogeneous, long ties reduce reinforcement for high-threshold individuals but create reach for low-threshold individuals. Across all seven empirically elicited threshold distributions, a fraction of low-threshold individual persists, and long ties do not hinder diffusion in our simulations.

Whether long ties can hinder diffusion in practice therefore depends on whether the adoption settings produce distributions our manipulations do not reach: those in which low-threshold adoption is rare and social reinforcement is required throughout the population \cite{eckles2024long}. Such distributions may occur for low-attractiveness or high-uncertainty products, coordination or normative behaviors whose payoffs depend on others' choices \cite{centola2018experimental}, settings where non-adoption is the default \cite{samuelson1988statusquo}, or situations where attention is limited \cite{sassine2024does}. Our results offer one possible reading of prior experimental work documenting a negative effect of long ties on health-behavior diffusion \cite{centola2010spread}. The setting in Ref.~\cite{centola2010spread} might generate a smaller share of low-threshold individuals because of several features: the behavior offers limited and uncertain payoff, the recruited population is relatively homogeneous, and non-adoption is the default. We cannot adjudicate this without direct threshold measurement. 

The implications of our results are expected to hold for adoption decisions that individuals weigh consciously rather than inheriting from a default option, and that cannot be postponed once the choice is presented. Here, individual differences widen the threshold distribution and leave low-threshold individuals for long ties to reach, so random rewiring raises adoption. The benefit of random rewiring narrows as decisions grow more uncertain or adoption less attractive. This suggests a complementarity between behavioral and structural interventions \cite{chater2022i-frame}: reducing uncertainty about a decision, or risk aversion in the population, may shift the threshold distribution and therefore affect the gains from rewiring interventions.

Several limitations bound these conclusions. The strategy method elicits contingent choice profiles, so temporal dynamics and normative pressure are absent from the design; it may also encourage monotonic responses, overstating the prevalence of clean threshold rules. The products are abstract monetary lotteries, so the results may not generalize to behaviors that are public, identity-relevant, or morally charged. Participants observed peer adoption counts but not network topology or adopter identity, so the manipulation operates through exposure quantity rather than perceived structural position or tie strength. In the simulations, thresholds are assigned to network positions randomly; real diffusion may differ if low-threshold individuals are clustered, peripheral, or disproportionately connected.

Our results suggest two directions for future research. First, identifying other factors that shift the threshold distribution, including normative pressure and identity relevance \cite{Turner2023uncertaintyform, centola2018experimental}, default structures that suppress active evaluation \cite{samuelson1988statusquo}, and temporal dynamics that allow thresholds to evolve as adoption unfolds \cite{sparkman2017dynamic}. Second, future research could use our empirical threshold distributions to study how network structure evolves with preferences and individual characteristics \cite{aral2009homophily, centola2011homophily, tur2023homophily}.

The threshold distribution is the natural bridge between behavioral research on decision making and theoretical research on social contagions. Our work contributes to a growing literature connecting these two traditions \cite{peres2010innovation, smith2017macro, bak-coleman2021stewardship, galesic-2021sensing, tanase2024integrating}, and suggests that questions about the effect of network structure are unlikely to be settled by either tradition alone.

\section*{Methods}

\begin{figure*}
\centering
\includegraphics[width=0.95\linewidth]{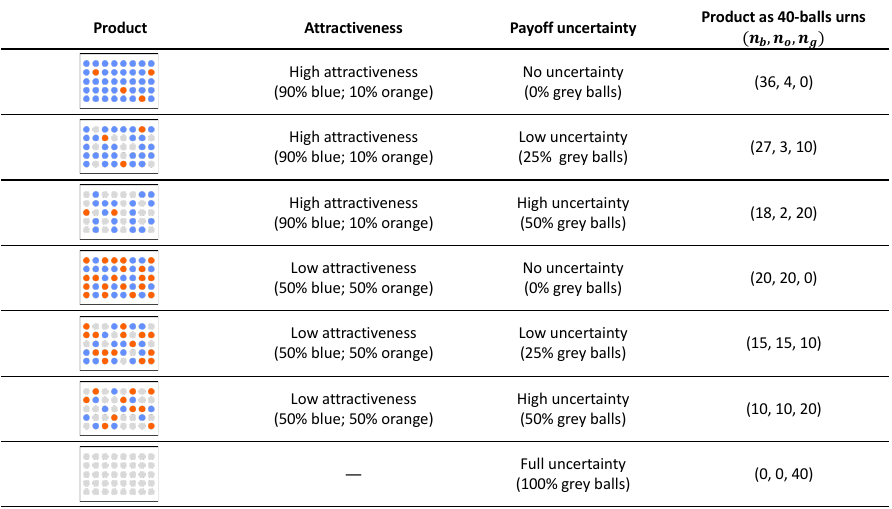}
\caption{\textbf{Product configurations.} Products are presented as urn lotteries. Each lottery is denoted by $(n_b, n_o, n_g)$ in an urn of $40$ balls: $n_b$ blue balls (giving a payoff of $300$), $n_o$ orange balls (giving payoff of $0$), and $n_g$ gray balls (hidden color). The seven products vary along two dimensions: payoff uncertainty, given by the proportion of hidden (gray) balls ($0\%$, $25\%$, $50\%$, $100\%$), and attractiveness, defined as the probability of a winning $300$ based on the visible balls (high attractiveness: $90\%$; low attractiveness: $50\%$).
}
\label{fig:lottery}
\end{figure*}

\subsection*{Participants and conditions}
The study was approved by the Institutional Review Board of the University of Zurich and preregistered (AsPredicted \#228275). We recruited $572$ participants from Prolific; 500 completed the study after passing mandatory comprehension checks covering the task, the payoff structure, and the threshold elicitation (13\% attrition). The sample comprised $200$ women, $295$ men and $1$ non-binary participant (4 did not report gender); mean age $41.7$ years (s.d.\ $13.1$). All participants provided informed consent. Of the 500, we randomly assigned $399$ to the social-information condition reported here and $101$ to a no-social-information condition, in which participants made the same decisions without peer information, providing a baseline estimate of intrinsic adoption propensity (see Supplementary Information, Sec.~\ref{si:baseline}).

\subsection*{Products as lotteries}
We represented products as Ellsberg-style urn lotteries \cite{ellsberg1961risk}. Each lottery $(n_b,n_o,n_g)$ comprised an urn of $40$ balls: $n_b$ blue (paying $300$), $n_o$ orange (paying $0$) and $n_g$ gray (hidden colour). Participants saw $(n_b,n_o,n_g)$ but not the composition of the gray balls, so the true success probability $\pi$ was fixed but unknown and bounded by $[L,U]$ with $L=n_b/40$ and $U=(n_b+n_g)/40$. Adopting a product paid $300$ points on a blue draw and $0$ otherwise; not adopting paid a sure $100$ points. We varied two product characteristics: payoff uncertainty, the proportion of gray balls, at four levels ($0\%$, $25\%$, $50\%$, $100\%$); and product attractiveness, the visible winning probability (the ratio of blue to orange among non-gray balls), at two levels ($90\%$ and $50\%$). Because the realized probability was unknown under uncertainty, attractiveness was defined from the visible composition alone. Under full uncertainty all balls were gray and the attractiveness manipulation was inoperative, yielding seven product configurations (Fig.~\ref{fig:lottery}; instructions in Supplementary Information, Fig.~\ref{fig:adoption-task}). Each participant evaluated all seven products in randomized order.

\subsection*{Threshold elicitation and incentives}
For each product we elicited an adoption threshold with the strategy method \cite{selten1967strategiemethode}: participants reported a separate adoption decision for each possible number of adopting peers ($0$--$4$), yielding a contingent choice profile per participant--product pair. After the experiment, each participant was assigned to a group of four others; one product configuration was drawn to be payoff-relevant (one attractiveness level applied to all participants, one uncertainty level drawn per participant), and the participant's recorded decision at the realized number of adopting peers determined their payoff. Because the realized count depended only on others' decisions, truthful reporting at every peer level maximized expected earnings, following the standard incentive logic of the strategy method \cite{selten1967strategiemethode, fischbacher-2001cooperation, charness-2002socialpreferences}. Recovering the realized count from contingent profiles required a resolution rule; the rule and a formal incentive-compatibility argument are given in Supplementary Information, Sec.~\ref{si:ic}.

\subsection*{Individual-difference measures}
We measured two participant-level covariates. Risk preferences were elicited with the multiple price list \cite{holt2002risk} (Supplementary Information, Fig.~\ref{fig:riskpreference}): across ten paired choices, participants chose between a safer Option~A and a riskier Option~B, with the probability of the high payoff increasing from $0.1$ to $1.0$; risk aversion was the proportion of safe (Option~A) choices. Subjective probability of success was each participant's stated probability of drawing a blue ball in the low-attractiveness ($50\%$ visible) products (Supplementary Information, Fig.~\ref{fig:estimation-task}); the regression used the estimate under high uncertainty ($50\%$ gray), the product with the widest range of admissible beliefs ($[0.25,0.75]$). The risk-preference task was incentivized; the belief elicitation was not.

\subsection*{Statistical analysis}
We constructed thresholds from the five binary decisions per participant--product pair. Throughout, $i$ indexes participants and $j$ product configurations. For each pair, let $m\in\{0,1,2,3,4\}$ denote the number of adopting peers shown (degree $k=4$ during elicitation). A response pattern was monotonic if adoption was non-decreasing in $m$; for monotonic patterns the threshold was the smallest $m$ at which the participant adopted, expressed in peer-count units ($0$--$4$; equivalently the fraction $m/4$), and participants who never adopted were assigned $\emptyset$ (unconditional non-adoption). Non-monotonic patterns ($7.1\%$ of observations) were excluded from threshold-based analyses but retained for the descriptive distributions (Fig.~\ref{fig:threshold-distributions}) and the simulations.

We modeled thresholds with a two-stage mixed-effects model, excluding the full-uncertainty product because attractiveness was not varied there (six configurations per participant). Stage~1 was a logistic mixed model of whether participant $i$ adopted configuration $j$ at any peer level ($\text{adopt}_{ij}=1$) versus never (unconditional non-adoption, $\emptyset$; $\text{adopt}_{ij}=0$); $N_{\text{obs}}=1{,}956$ over $N_{\text{participants}}=326$. It had intercept $\beta_0$, fixed-effect coefficient vector $\boldsymbol{\beta}$ acting on the predictor vector $\mathbf{X}_{ij}$, and a participant random intercept $u_i$:

\begin{align*}
\text{logit}\bigl(\Pr(\text{adopt}_{ij}=1)\bigr)
&= \beta_0 + \mathbf{X}_{ij}\boldsymbol{\beta} + u_i, \\
&\quad u_i \sim \mathcal{N}(0, \tau_u^2).
\end{align*}

Stage~2 was a linear mixed model of the adoption threshold (in peer-count units, $0$--$4$) among adopters ($N_{\text{obs}}=1{,}670$ over $N_{\text{participants}}=320$; the six participants who never adopted contributed to Stage~1 only), with intercept $\gamma_0$, fixed-effect coefficients $\boldsymbol{\gamma}$, a participant random intercept $v_i$, and residual $\varepsilon_{ij}$:

\begin{align*}
\text{threshold}_{ij}
&= \gamma_0 + \mathbf{X}_{ij}\boldsymbol{\gamma} + v_i + \varepsilon_{ij}, \\
&\quad v_i \sim \mathcal{N}(0, \tau_v^2), \\
&\quad \varepsilon_{ij} \sim \mathcal{N}(0, \sigma^2).
\end{align*}

Here $\tau_u^2$ and $\tau_v^2$ are the participant-level (random-intercept) variances in Stages~1 and~2, and $\sigma^2$ the Stage~2 residual variance. The predictor vector $\mathbf{X}_{ij}$ comprised product attractiveness (binary), payoff uncertainty (proportion of hidden balls), risk aversion, subjective probability of success, age, education and gender; continuous predictors were standardized (mean $0$, s.d.\ $1$). 

Average marginal effects were averaged over the empirical covariate distribution with delta-method $95\%$ confidence intervals, and marginal and conditional $R^2$ followed Nakagawa et al.\ \cite{nakagawa2017coefficient}. Models were fitted in R~4.5.2 with \texttt{lme4} \cite{bates2015fitting}, \texttt{lmerTest} \cite{kuznetsova2017lmertest}, \texttt{marginaleffects} \cite{arel-bundock2024marginaleffects} and \texttt{performance} \cite{ludecke2021performance}.

\subsection*{Network simulations}
We embedded the elicited thresholds in an agent-based susceptible--infected model. The main simulations used ring lattices of $N=399$ nodes, matching the sample, with degree $k=4$, matching the four peers shown during elicitation so that thresholds were evaluated at the levels at which they had been measured. We introduced long ties by degree-preserving rewiring \cite{maslov2002specificity}, from $0$ to $200$ rewired edge pairs. To assess robustness to realistic topology, we repeated the simulations on empirical social networks from three sources: Add Health school friendships \cite{guilbeault2021topological}, Chinese villages \cite{cai2015social} and Indian villages \cite{banerjee2023changes}. We retained the largest connected component of each and comparing the original network with one in which $10\%$ of edges were rewired (see Supplementary Information, Sec.~\ref{si:sim-empirical-selection}). The rewiring algorithm and the rule for evaluating thresholds when $k\neq4$ are given in Supplementary Information, Secs.~\ref{si:sim-rewiring} and~\ref{si:sim-offgrid}.

In each run we drew $N$ thresholds with replacement from the sample and assigned one to each node, so that variability across runs reflected sampling uncertainty in the threshold distribution alongside network and seed randomization. Diffusion was seeded from one randomly chosen connected pair of nodes, set to adopted; a connected pair supplied the local reinforcement that thresholds above $1/4$ required. At each step, every non-adopter with at least one adopting neighbor evaluated its threshold at the current number of adopting neighbors and updated synchronously, so that decisions at time $t$ depended only on the state at $t-1$ \cite{centola2007complex, watts-2002simplemodel}. Since evaluation began only at the first adopting neighbor, thresholds $0$ and $1/4$ were dynamically indistinguishable. Runs continued until no further adoptions occurred or $100$ steps elapsed. For each product, network and rewiring level we ran $R=500$ independent simulations, each with a fresh threshold draw, network and seed; the outcome was the final fraction of adopters. Robustness to degree, network size and seed size is reported in Supplementary Information, Figs.~\ref{fig:rob-degree}, ~\ref{fig:rob-netsize} and~\ref{fig:rob-seed}.

\subsection*{Code and data availability}
Anonymized data and analysis code will be deposited in a public repository on acceptance.

\subsection*{Author contributions}
L.L., M.S.M., R.A., R.T. designed research; L.L. performed research; L.L. analyzed data; L.L. wrote the first draft of the paper; all authors revised the paper.

\subsection*{Acknowledgments}
We thank P. Smaldino, E. Fehr, attendees in the Quantitative Marketing Research seminar at the University of Zurich and ETH Zurich, as well as attendees at the Sunbelt and IC2S2 conferences, for their helpful comments and insightful discussions. This work has been supported by the URPP Social Networks and the Swiss National Science Foundation (SNSF) (Grant No. 100013--207888 and 100013--236802).

\section*{References}
\bibliography{preprint}

\clearpage
\onecolumngrid

\setcounter{section}{0}
\setcounter{figure}{0}
\setcounter{table}{0}
\renewcommand{\thesection}{S\arabic{section}}
\renewcommand{\thefigure}{S\arabic{figure}}
\renewcommand{\thetable}{S\arabic{table}}
\renewcommand{\theHsection}{supplementary.S\arabic{section}}
\renewcommand{\theHfigure}{supplementary.S\arabic{figure}}
\renewcommand{\theHtable}{supplementary.S\arabic{table}}
\setcounter{secnumdepth}{2}
\setcounter{tocdepth}{2}

\begin{center}
    {\large\bf Supplementary Information for:}\\[4pt]
    {\bf What shifts threshold distributions in social contagions?}\\[6pt]
    Luca Lazzaro, Manuel S. Mariani, Ren\'e Algesheimer, Radu Tanase
\end{center}

\medskip
\tableofcontents
\clearpage

\section{Experimental design and stimuli}\label{si:design}
\begin{figure}[h]
\centering
\includegraphics[width=.7\linewidth]{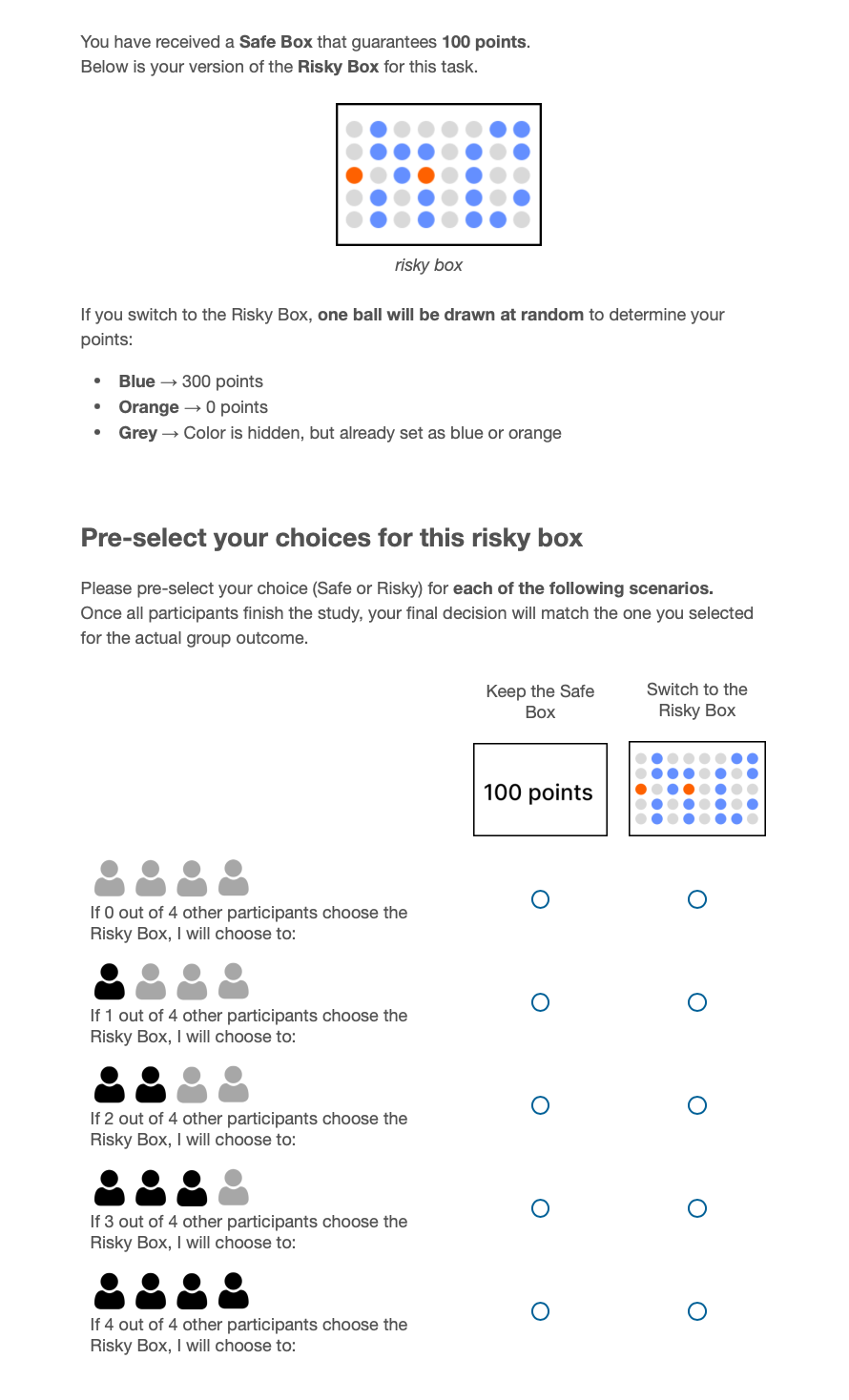}
\caption{\textbf{Adoption task eliciting threshold.} Screenshot of the interface participants saw: for one product, they pre-specified whether they would adopt at each possible number of adopting peers (a contingent choice profile over $k\in\{0,\dots,4\}$). This adapts the strategy method to elicit adoption thresholds \cite{selten1967strategiemethode}. The screenshot shows the high-attractiveness, high-uncertainty product.}
\label{fig:adoption-task}
\end{figure}

\begin{figure}[h]
\centering
\includegraphics[width=.75\linewidth]{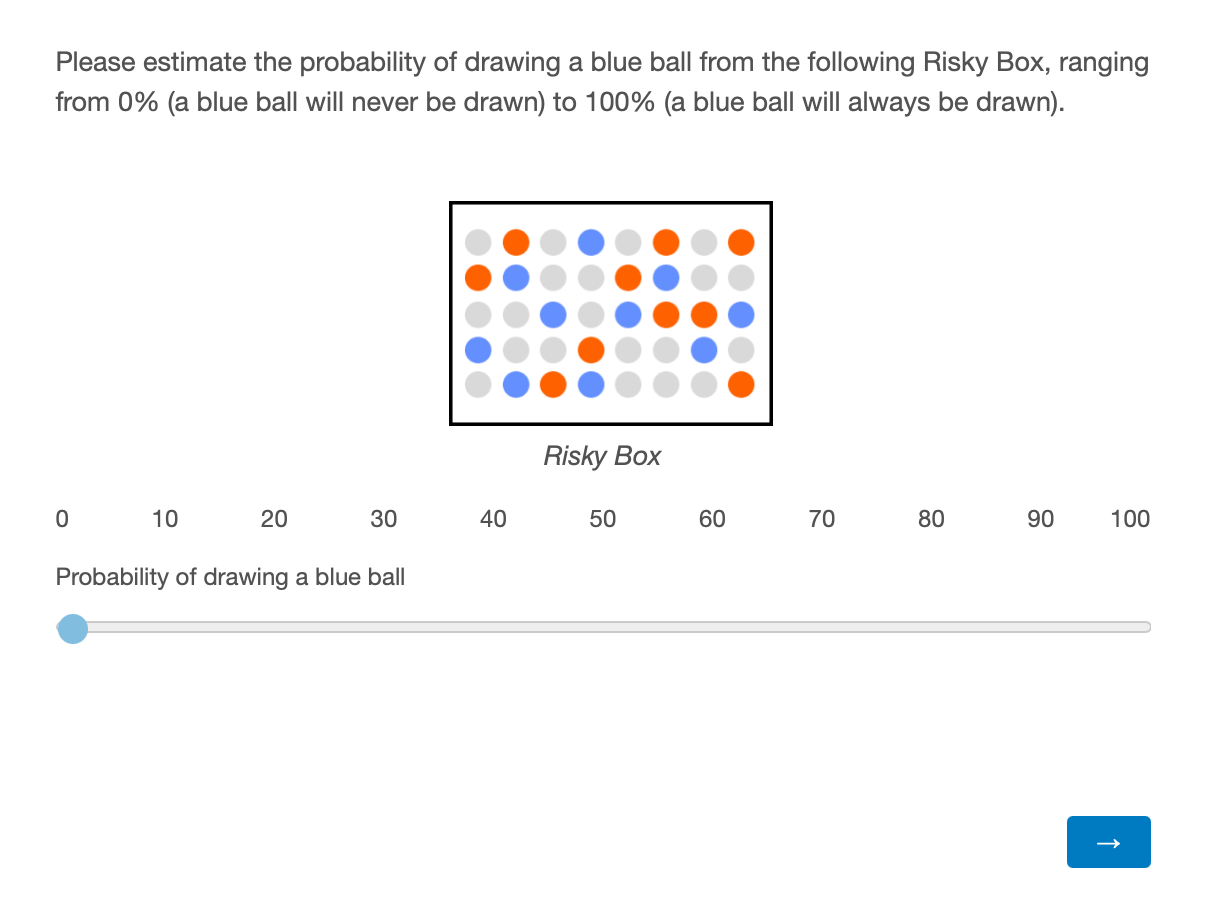}
\caption{\textbf{Probability estimation task.} Screenshot of the interface participants saw: they estimated the probability of drawing a blue ball from a 40-ball urn under low attractiveness (50\% blue, 50\% orange) at three uncertainty levels: no (0\% gray balls), low (25\%), and high (50\%). The screenshot shows the low-attractiveness, high-uncertainty product. We use each participant's estimate for the high-uncertainty product as the subjective probability of success, which we use to proxy how individuals resolve uncertainty.}
\label{fig:estimation-task}
\end{figure}

\begin{figure}[h]
\centering
\includegraphics[width=.95\linewidth]{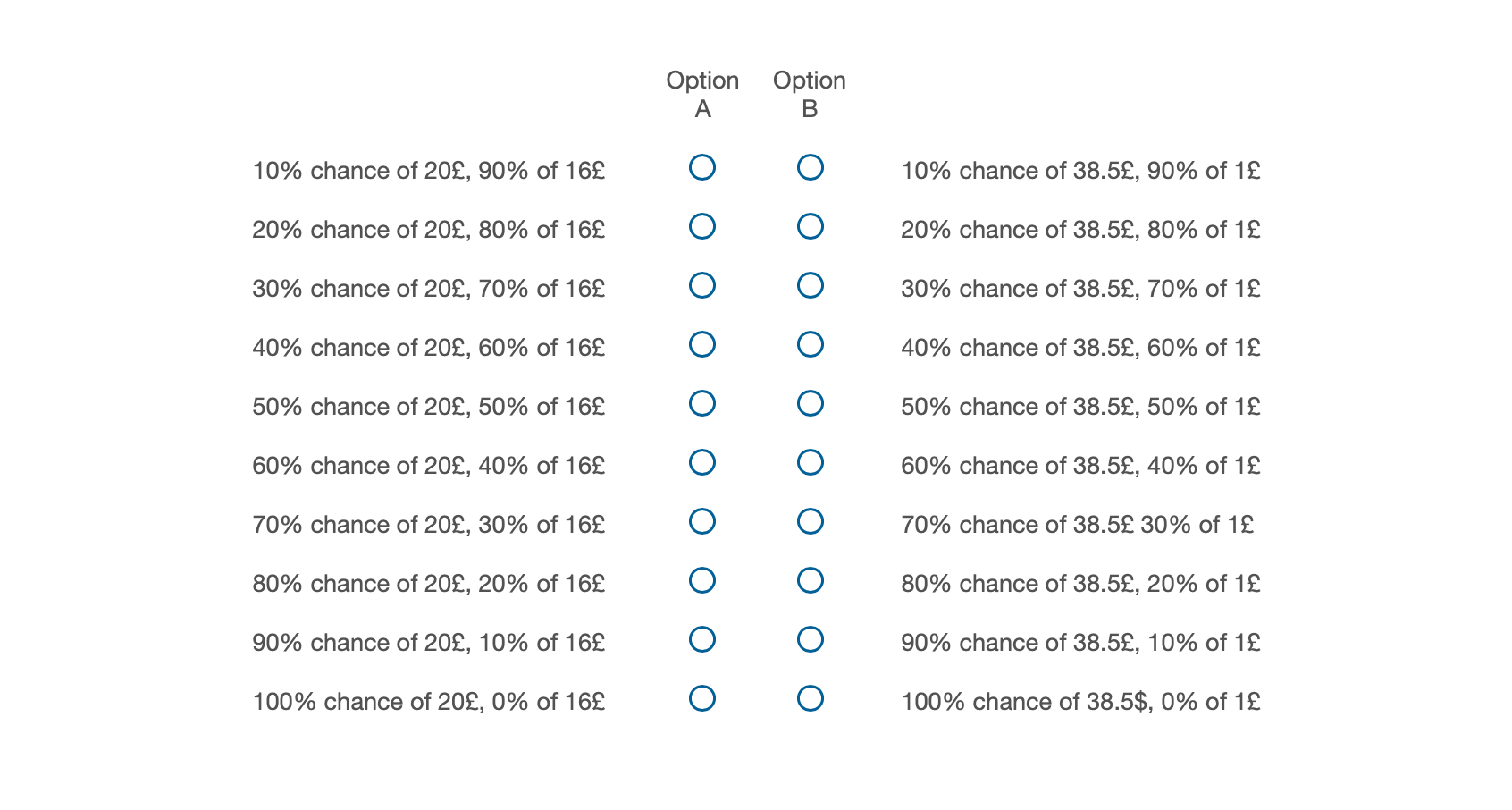}
\caption{\textbf{Risk preference task.} Screenshot of the interface participants saw: choices between paired lotteries used to elicit risk preferences with the multiple price list method \cite{holt2002risk}. Option A is the relatively safe lottery (smaller payoff variance); Option B is the riskier lottery (higher variance). The probability of the high payoff increases in increments of 0.1 across the ten rows. Risk aversion is the proportion of safe (Option A) choices across the ten decisions.}
\label{fig:riskpreference}
\end{figure}

This section reproduces the on-screen stimuli for the three elicitation tasks described in Methods: the adoption task used to elicit thresholds (Fig.~\ref{fig:adoption-task}), the probability-estimation task used to measure subjective probability of success (Fig.~\ref{fig:estimation-task}), and the multiple-price-list task used to measure risk preferences (Fig.~\ref{fig:riskpreference}). Each screenshot shows one representative product configuration; the full set of seven configurations is given in Fig.~\ref{fig:lottery}.

\subsection{Adoption task}\label{si:design-adoption}

\clearpage

\subsection{Probability-estimation task}\label{si:design-estimation}

\clearpage

\subsection{Risk-preference task}\label{si:design-risk}

\clearpage

\section{Threshold elicitation and incentive compatibility}\label{si:ic}
Here we show that truthful reporting is optimal at every number of adopting peers, so the strategy method recovers each participant's threshold.

The strategy method elicits each participant's threshold as a contingent choice profile, an adoption decision at every possible number of adopting peers \cite{selten1967strategiemethode}. Participants were told that, after the experiment, they would be assigned to a group of four other participants; that the number of group members who chose to adopt would be counted; and that their recorded decision at that count would determine their payoff. Because the realized count depended only on the other participants' decisions, which the focal participant could neither observe nor influence at the time of reporting, any of the five rows of the contingent choice profile could become payoff-relevant. Misreporting at any row therefore produced a strict expected loss in the states where that row was realized, with no offsetting gain elsewhere. Crucially, participants were instructed that this count came from the other group members' decisions, which they could neither observe nor influence when reporting: their own entries selected which decision was evaluated, but not which state was realized.

The resolution rule used to realize payoffs placed all $399$ participants on a random network with fixed degree $k = 4$ and applied their elicited profiles synchronously until no further adoptions occurred; each participant's payoff was determined by their decision at the realized number of adopting neighbors at convergence. The choice of resolution rule does not affect the elicited threshold, which is a primitive of the participant's adoption decision under the elicited information: the participant knew the product configuration and the number of adopting peers, but not peer identities, network positions, or temporal trajectories.

\clearpage

\section{Threshold summary statistics}\label{si:tsummary}

\begin{figure}[h]
\centering
\includegraphics[width=1\linewidth]{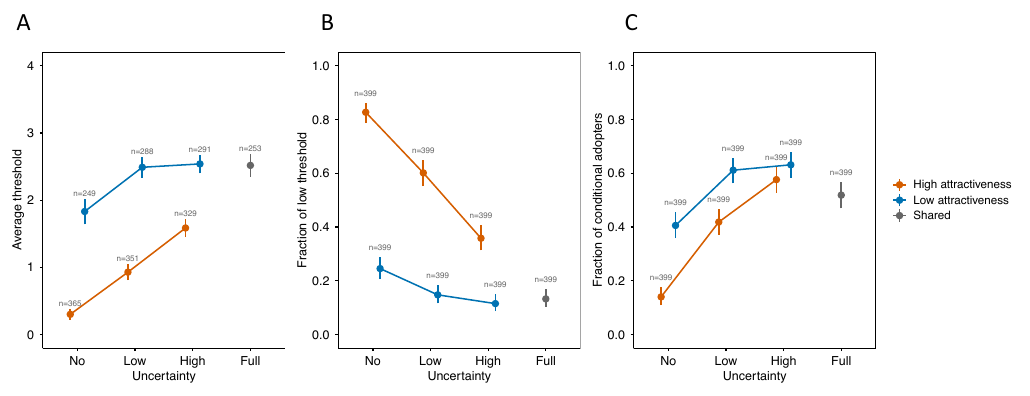}
\caption{\textbf{Threshold summary statistics across product configurations.}  Each panel summarizes adoption thresholds across the seven product configurations---the $2\times3$ (attractiveness: high, low; uncertainty: no, low, high) design plus a shared full-uncertainty condition. (A) Average adoption threshold among participants with monotonic response patterns (adopters), the average number of adopting peers required before adoption (range $0$--$4$). (B) Fraction of participants with low thresholds ($0$ or $1/4$). (C) Fraction of conditional adopters (threshold of $1/4$, $2/4$, $3/4$, or $4/4$)---those who adopt but require at least some peer adoption. Across all three measures, lower attractiveness and higher uncertainty shift the distribution toward higher thresholds: average thresholds increase, the fraction of low thresholds thresholds declines, and the fraction of conditional adopters rises. Under full uncertainty, the conditional-adopter fraction declines, reflecting a compositional shift from conditional adoption to unconditional non-adoption. Error bars in (A) are $95$\% $t$-intervals for the mean; in (B) and (C) they are $95$\% Wilson intervals for the proportion. Sample sizes are annotated above each point.}
\label{fig:t-summary}
\end{figure}

\clearpage

\section{Full two-stage mixed-effects model}\label{si:fullmodel}
Table~\ref{tab:hurdle-main} reports both stages of the mixed-effects model side by side: Stage~1 (adoption, odds ratios) and Stage~2 (threshold among adopters, $\gamma$ in peer-count units). Because the two stages model different quantities, a predictor may enter with opposite signs across stages, and the two columns are not on a common scale.

\begin{table} [h]
\centering
\caption{\textbf{Full results of the two-stage mixed-effects model.} Stage~1 models adoption (binary logit); Stage~2 models the adoption threshold among adopters (linear mixed model; outcome: $0$--$4$ adopting peers). Both stages include participant random intercepts. Predictors are standardized. Stage~1 reports odds ratios (OR); Stage~2 reports linear coefficients ($\gamma$). $95$\%
confidence intervals and $p$-values are reported. The full-uncertainty condition is excluded (attractiveness not manipulated).}
\label{tab:hurdle-main}
\begin{ruledtabular}
\begin{tabular}{lrrrrrr}
& \multicolumn{3}{c}{\textbf{Stage 1: Adoption (logit)}} & \multicolumn{3}{c}{\textbf{Stage 2: Threshold (linear)}} \\
\cline{2-4}\cline{5-7}
Predictor & OR & 95\% CI & p-value & $\gamma$ & 95\% CI & p-value \\
\hline
Payoff uncertainty      & 1.29  & [1.09, 1.54]    & 0.003     & 0.43    & [0.38, 0.48]           & $<$0.001 \\
Attractiveness (high vs.\ low)  & 16.46 & [10.08, 26.84]  & $<$0.001  & $-$1.43 & [$-$1.52, $-$1.34]     & $<$0.001 \\
Risk aversion           & 0.43  & [0.31, 0.62]    & $<$0.001  & 0.08    & [$-$0.02, 0.17]        & 0.109 \\
Subjective probability of success      & 1.55  & [1.12, 2.15]    & 0.008     & $-$0.16 & [$-$0.26, $-$0.07]     & 0.001 \\
Age                     & 0.86  & [0.62, 1.20]    & 0.375     & 0.03    & [$-$0.07, 0.12]        & 0.582 \\
Education               & 0.87  & [0.63, 1.20]    & 0.400     & 0.07    & [$-$0.03, 0.16]        & 0.151 \\
Gender (male)           & 1.26  & [0.65, 2.44]    & 0.488     & $-$0.22 & [$-$0.41, $-$0.03]     & 0.027 \\
\hline
\multicolumn{7}{l}{\textbf{Random effects}} \\
Residual variance ($\sigma^2$) & -- & & & 0.86 & & \\
Intercept variance ($\tau_u^2$ / $\tau_v^2$)             & 4.94 & & & 0.54 & & \\
ICC                              & 0.60 & & & 0.39 & & \\
N Participants                       & 326  & & & 320  & & \\
\hline
\multicolumn{7}{l}{\textbf{Model statistics}} \\
Observations          & 1{,}956 & & & 1{,}670 & & \\
Marginal $R^2$        & 0.269   & & & 0.349   & & \\
Conditional $R^2$     & 0.711   & & & 0.603   & & \\
\end{tabular}
\end{ruledtabular}
\end{table}

\clearpage

\section{Network simulations: additional details}\label{si:simdetails}
This section provides the simulation details deferred from Methods:  the degree-preserving rewiring algorithm used to introduce long ties (Sec.~\ref{si:sim-rewiring}), the sources and inclusion criteria for the empirical networks (Sec.~\ref{si:sim-empirical-selection}), and the rule for evaluating elicited thresholds when network degree differs from the elicitation grid (Sec.~\ref{si:sim-offgrid}). The seeding and diffusion procedure applies uniformly to synthetic and empirical networks and is described in Methods. 

\subsection{Rewiring algorithm}\label{si:sim-rewiring}

We introduce long ties by progressively rewiring pairs of edges with a degree-preserving double-edge swap \cite{maslov2002specificity}. Each swap (i) selects two edges uniformly at random; (ii) proposes exchanging their endpoints; (iii) rejects the proposal if it would create a self-loop or a multi-edge; (iv) rejects it if it would disconnect the graph; and (v) repeats until a valid swap is found. The procedure preserves the degree of every node and the connectedness of the network, so any change in diffusion is attributable to the rewiring of ties rather than to a change in the degree sequence.

For the synthetic ring lattices, the number of swaps ranges from $0$ (the original ring lattice) to $200$ rewired edge pairs. For the empirical networks, we compare each original network ($0\%$ rewired) with a counterfactual in which $10\%$ of its edges are rewired, where the percentage denotes the fraction of edges that undergo a successful swap.

\clearpage

\subsection{Empirical networks: sources and selection criteria}\label{si:sim-empirical-selection}

To assess robustness to realistic network topology, we replicate the main simulations on empirical social networks from three independent sources: Add Health school-friendship networks \cite{guilbeault2021topological}, village networks in China \cite{cai2015social}, and village networks in India \cite{banerjee2023changes}. These networks have heterogeneous degree distributions and community structure absent from ring lattices. We selected these datasets because they record genuine social ties between individuals; their size and average degree fall within the range relevant to product diffusion; and each dataset contains many independent networks from a distinct population, letting us assess variability across networks of similar type and across social contexts. For each network we retain the largest connected component and exclude networks with fewer than six edges. This leaves $76$ Add Health networks, $153$ Chinese village networks, and $75$ Indian village networks. Table~\ref{tab:emp-descriptives} presents the descriptive of the networks.

\begin{table}[h]
\centering
\caption{\textbf{Empirical social networks used in the diffusion simulations.} For each source we retain the largest connected component of every network and exclude networks with fewer than six edges (for which a $10\%$ rewiring rounds to zero swaps). Mean clustering is averaged across the included networks, before ($0\%$) and after ($10\%$) degree-preserving rewiring; the reduction confirms that rewiring removes local clustering and introduces long ties.}
\begin{ruledtabular}
\begin{tabular}{lrrrrr}
 & Networks & Size range & Mean degree & Clustering ($0\%$) & Clustering ($10\%$) \\
\hline
Add Health \citep{guilbeault2021topological}         & 76  & 25--2539 & 7.30  & 0.236 & 0.140 \\
Cai et al.\ (2015) \citep{cai2015social}             & 153 & 5--95    & 5.20  & 0.291 & 0.250 \\
Banerjee et al.\ (2024) \citep{banerjee2023changes}  & 75  & 77--382 & 17.67 & 0.314 & 	0.222 \\
\end{tabular}
\end{ruledtabular}
\label{tab:emp-descriptives}
\end{table}

\clearpage

\subsection{Off-grid threshold evaluation}\label{si:sim-offgrid}

Adoption thresholds are elicited at five signal levels, $s\in\{0, 1/4, 2/4, 3/4, 1\}$, corresponding to the fraction of adopting neighbors in the experimental network (degree $k=4$). In the synthetic main simulations ($k=4$) every realized signal falls on this grid, so a participant's elicited decision applies directly. In the robustness simulations and empirical networks, where $k\neq4$, the realized signal $s=m/k$ (with $m$ adopting neighbors) generally falls between two elicited levels, and the elicited profile does not specify a decision at that point.

We resolve this by linear interpolation between the two bracketing levels. Let $d_0$ and $d_1$ be the participant's elicited decisions at the grid points immediately below and above $s$, and let $\alpha\in[0,1]$ be the fractional distance of $s$ across that interval. The agent adopts with probability $p = d_0 + \alpha\,(d_1 - d_0)$, and the adoption draw is $\mathrm{Bernoulli}(p)$. The rule uses only the two responses local to the realized signal, imposing no assumption on the shape of the response function elsewhere.

For example, consider a node of degree $k=8$ with $m=3$ adopting neighbors. The realized signal is $s = 3/8 = 0.375$, which lies between the elicited grid points $1/4$ and $2/4$ at fractional distance $\alpha = (0.375-0.25)/(0.5-0.25) = 0.5$. A participant who adopts at $2/4$ but not at $1/4$ ($d_0=0$, $d_1=1$) therefore adopts with probability $p = 0.5$; a participant who adopts at both levels ($d_0=d_1=1$) adopts with certainty. When $s$ falls exactly on a grid point the rule reduces to the elicited decision and is deterministic, so the main results are unaffected.

\begin{figure}[h]
\centering
\includegraphics[width=.8\linewidth]{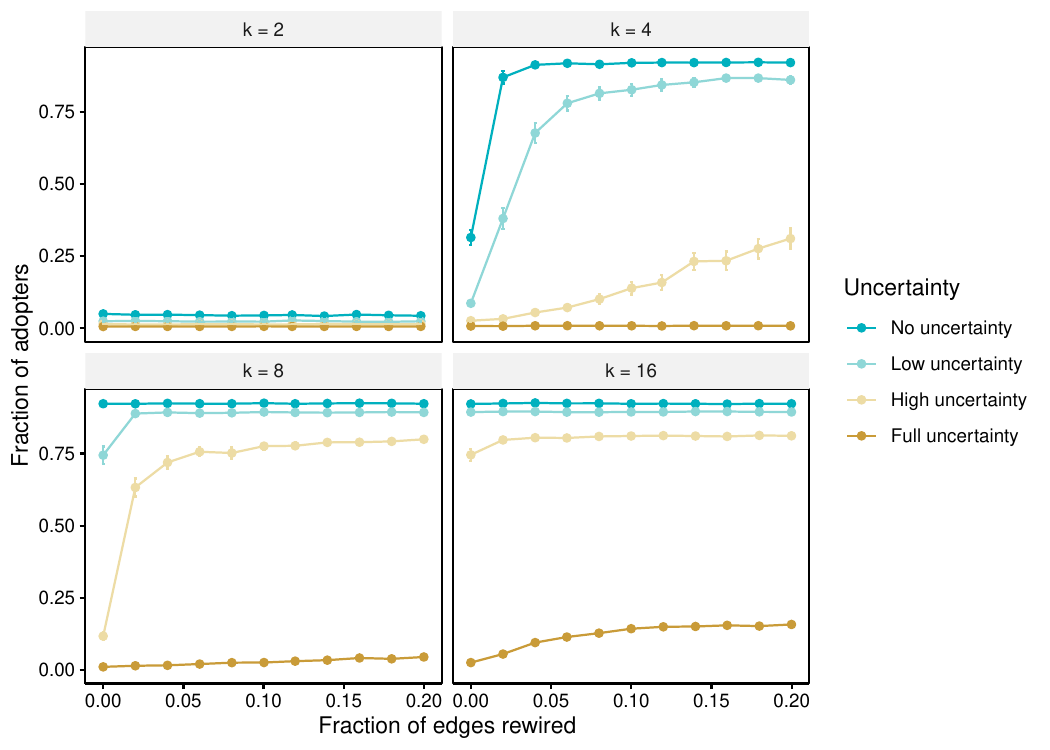}
\caption{\textbf{Robustness check: degree.} Average final fraction of adopters as a function of the fraction of edges rewired, separately for network degrees $k\in\{2,4,8,16\}$; the fractional scale places different degrees on a common $x$-axis. High-attractiveness products only; the low-attractiveness case is omitted because diffusion remains negligible at all rewiring levels. For $k\neq4$, realized signals $s=m/k$ are evaluated off-grid (Sec.~\ref{si:sim-offgrid}). Simulations use ring lattices ($N=399$) with degree-preserving rewiring (Sec.~\ref{si:sim-rewiring}); diffusion is seeded from a randomly chosen connected pair, and thresholds are resampled with replacement and randomly assigned to nodes each run. Long ties raise adoption at every degree, with the effect strengthening as degree increases (see text). Points are means across $R=200$ realizations; error bars are $95\%$ confidence intervals for the mean.}
\label{fig:rob-degree}
\end{figure}

Evaluating the threshold at the fraction $s=m/k$ holds the fractional threshold distribution fixed as degree varies: a given threshold corresponds to the same fraction of neighbors at every degree, but to a larger absolute number of adopting neighbors as $k$ grows. Because the experiment measures thresholds only at $k=4$, it cannot distinguish fractional thresholds from absolute (count-based) ones. The degree-robustness (Fig.~\ref{fig:rob-degree}) and empirical networks (Fig.~\ref{fig:rob-empirical}) results should be read under the fractional interpretation this rule assumes.
\clearpage

\section{Additional simulation results and robustness}\label{si:robust}

\subsection{Long ties across risk-aversion populations}\label{si:rob-pop}
\begin{figure}[h]
\centering
\includegraphics[width=.6\linewidth]{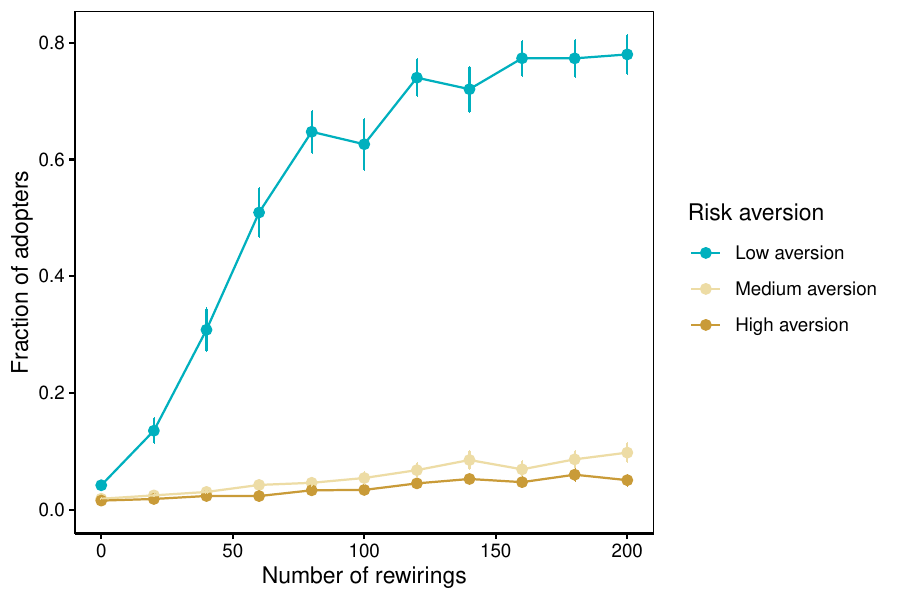}
\caption{\textbf{Effect of long ties across populations differing in risk aversion.} Average final fraction of adopters as a function of the number of rewired edge pairs, for a fixed product (high attractiveness, high uncertainty) and three synthetic populations defined by risk-aversion tercile. Participants ($N=399$) are split into terciles by their multiple-price-list score \cite{holt2002risk} (cutpoints computed on the full social-influence sample): low ($n=180$), medium ($n=131$), and high ($n=88$) risk aversion. For each realization, $N=399$ thresholds are drawn with replacement from the relevant tercile and assigned to nodes at random, so the resulting threshold distribution reflects that tercile's risk composition. Simulations use a ring lattice ($N=399$, $k=4$) with degree-preserving rewiring (Sec.~\ref{si:sim-rewiring}). Points are means across $R=200$ realizations; error bars are $95\%$ confidence intervals for the mean.}
\label{fig:pop-comp}
\end{figure}

Holding the product fixed, we ask whether the effect of long ties varies with the population's risk aversion composition. We resample thresholds within risk-aversion terciles to build synthetic populations that differ in risk aversion, and run the main diffusion simulations on each. Consistent with the across-product result in the main text (Fig.~\ref{fig:macro_results}), less risk-averse populations, which contain more low-threshold individuals, show larger gains from long ties, and more risk-averse populations show weaker gains (Fig.~\ref{fig:pop-comp}). The effect of long ties therefore depends on the product and on the population composition.

\clearpage

\subsection{Magnitude of the effect of long ties}\label{si:rob-magnitude}
The main text shows that long ties benefit diffusion but that the size of the benefit varies with the threshold distribution. Here we quantify that variation with a single summary: the \emph{total gain}, the increase in the final fraction of adopters when the network goes from a pure ring lattice to fully rewired. A large total gain means long ties add many adopters; a total gain near zero means they add no.

\begin{figure}[h]
\centering
\includegraphics[width=.8\linewidth]{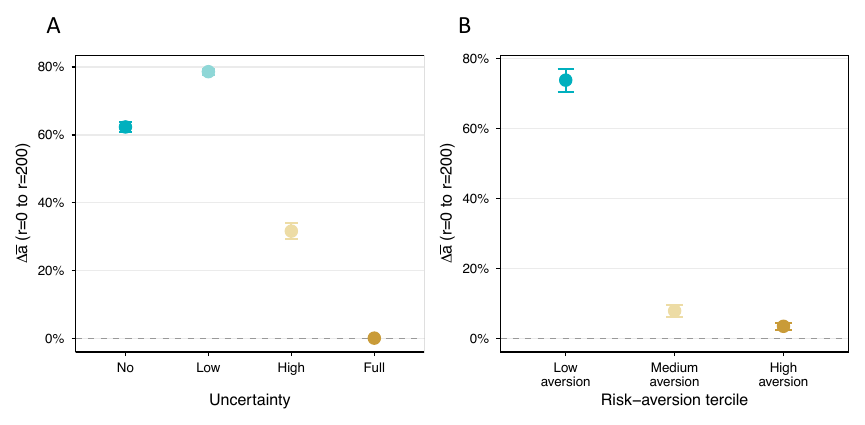}
\caption{\textbf{Magnitude of the effect of long ties across uncertainty levels and risk-aversion populations.} Total gain, the difference in the final fraction of adopters between the fully rewired network and the pure ring lattice, with point estimates and $95\%$ confidence intervals across simulation runs. (A) High-attractiveness product across four payoff-uncertainty levels (no, low, high, full); $n=500$ runs per condition. (B) Three risk-aversion terciles (low, medium, high), product fixed at high attractiveness and high uncertainty; $n=200$ runs per tercile. Simulations use ring lattices ($N=399$, $k=4$) with degree-preserving rewiring (Sec.~\ref{si:sim-rewiring}).}
\label{fig:magnitude}
\end{figure}

\paragraph*{Across uncertainty.} Fixing a high-attractiveness product and varying payoff uncertainty, the total gain is non-monotone (Fig.~\ref{fig:magnitude}A). It peaks at low uncertainty ($0.79$), is smaller with no uncertainty ($0.62$), and falls toward zero under high ($0.32$) and full ($0.00$) uncertainty. The peak sits at an intermediate distribution for an intuitive reason: with no uncertainty the population splits sharply into unconditional adopters and non-adopters, leaving few contingent individuals for long ties to tip; low uncertainty moves more of the population into the contingent range, where long ties have the most to work with; and high uncertainty pushes thresholds too high for long ties to overcome.

\paragraph*{Across risk aversion.} Fixing the product and varying the population's risk composition, the total gain declines monotonically with risk aversion (Fig.~\ref{fig:magnitude}B): it is large for the least risk-averse population ($0.74$) and collapses for the medium ($0.08$) and most risk-averse ($0.04$) populations, whose high thresholds long ties cannot satisfy.

\clearpage

\subsection{Empirical social networks}\label{si:rob-empirical}
\begin{figure}[h]
\centering
\includegraphics[width=1\linewidth]{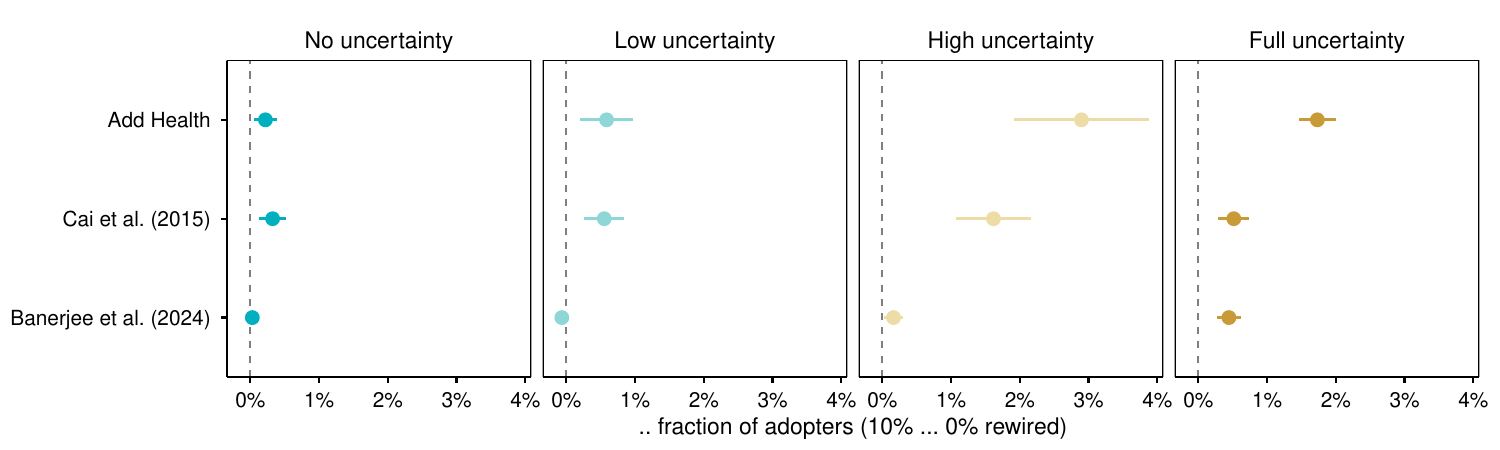}
\caption{\textbf{The effect of long ties on diffusion across empirical social networks.} Effect size $\Delta$, the change in the final fraction of adopters when $10\%$ of a network's edges are rewired ($\Delta=$ adoption at $10\%$ minus at $0\%$), across the four payoff-uncertainty conditions, separately for each network source (Table~\ref{tab:emp-descriptives}). High-attractiveness products only; the low-attractiveness case is omitted because diffusion remains negligible at all rewiring levels. Because empirical degrees differ from four, realized signals $s=m/k$ are evaluated off-grid (Sec.~\ref{si:sim-offgrid}). Simulations use each network's largest connected component with degree-preserving rewiring (Sec.~\ref{si:sim-rewiring}); thresholds are resampled with replacement and randomly assigned to nodes each run. Long ties raise adoption in every condition, with the gain largest where baseline adoption is intermediate (see text). Points are means across networks, each averaged over $R=100$ realizations; error bars are $95\%$ confidence intervals across networks.}
\label{fig:rob-empirical}
\end{figure}

We replicate the long-ties analysis on empirical social networks (Table~\ref{tab:emp-descriptives}; sources and selection criteria in Sec.~\ref{si:sim-empirical-selection}), comparing final adoption before and after rewiring $10\%$ of edges. Table~\ref{tab:rob-empirical} reports the underlying adoption levels for high-attractiveness products; Fig.~\ref{fig:rob-empirical} shows the corresponding effect sizes $\Delta$ with confidence intervals across networks. Long ties do not decrease adoption in every condition, so on realistic topologies with our measured threshold distributions long ties never hinder diffusion.

The size of the effect follows the same intermediate-peak logic as the main text (Fig.~\ref{fig:magnitude}). Unlike the ring lattice, which contains no long ties, empirical networks already contain some and diffuse close to ceiling without any rewiring under no and low uncertainty (baseline adoption $0.86$--$0.92$). With almost no non-adopters left to convert, a $10\%$ rewiring moves adoption by at most half a percentage point. High uncertainty lowers baseline adoption into an intermediate range ($0.69$--$0.73$ for Add Health and Cai et al.\ (2015)), where long ties can expose more low-threshold adopters, producing the largest gains ($+2.9$ and $+1.6$~pp). Under full uncertainty, baseline adoption is low and thresholds are high enough that diffusion stalls, so the gain recedes. 

The peak therefore sits at higher uncertainty than on the ring lattice. The reason is the shape of the rewiring curve in each condition (Fig.~\ref{fig:macro_results}, main text). Under no and low uncertainty the effect peaks at low rewiring and is exhausted by the first few long ties, which empirical networks already possess. Under high uncertainty the gains accumulate steadily with rewiring, so the long ties already present in the empirical networks leave room for long ties to still move adoption. 

\begin{table}[h]
\centering
\caption{\textbf{Adoption levels across empirical social networks (high-attractiveness products).} Final fraction of adopters (\%) before ($0\%$) and after ($10\%$) degree-preserving rewiring, and their difference $\Delta$ (percentage points), by dataset and payoff-uncertainty condition. Values are means across networks and diffusion realizations; $95\%$ confidence intervals on $\Delta$ are $t$-intervals across networks.}
\begin{ruledtabular}
\begin{tabular}{llrrrr}
Dataset & Uncertainty & $0\%$ & $10\%$ & $\Delta$ (pp) & $95\%$ CI \\
\hline
Add Health \cite{guilbeault2021topological} & No & 91.5 & 91.7 & $+0.22$ & [$+0.05$, $+0.38$] \\
                                            & Low & 87.3 & 87.9 & $+0.58$ & [$+0.20$, $+0.97$] \\
                                            & High & 72.9 & 75.8 & $+2.90$ & [$+1.91$, $+3.88$] \\
                                            & Full & \phantom{0}5.4 & \phantom{0}7.1 & $+1.73$ & [$+1.46$, $+2.00$] \\
\hline
Cai et al.\ (2015) \cite{cai2015social} & No & 90.7 & 91.1 & $+0.32$ & [$+0.12$, $+0.52$] \\
                                        & Low & 85.9 & 86.4 & $+0.55$ & [$+0.26$, $+0.84$] \\
                                        & High & 69.2 & 70.8 & $+1.62$ & [$+1.07$, $+2.16$] \\
                                        & Full & 16.6 & 17.1 & $+0.52$ & [$+0.29$, $+0.74$] \\
\hline
Banerjee et al.\ (2024) \cite{banerjee2023changes} & No & 92.2 & 92.3 & $+0.03$ & [$-0.04$, $+0.09$] \\
                                                   & Low & 89.4 & 89.3 & $-0.07$ & [$-0.14$, $+0.01$] \\
                                                   & High & 81.0 & 81.2 & $+0.16$ & [$+0.03$, $+0.30$] \\
                                                   & Full & 15.7 & 16.2 & $+0.44$ & [$+0.27$, $+0.62$] \\
\end{tabular}
\end{ruledtabular}
\label{tab:rob-empirical}
\end{table}

\clearpage

\subsection{Robustness to network degree}\label{si:rob-degree}

\clearpage

\subsection{Robustness to network size}\label{si:rob-size}
\begin{figure}[h]
\centering
\includegraphics[width=.8\linewidth]{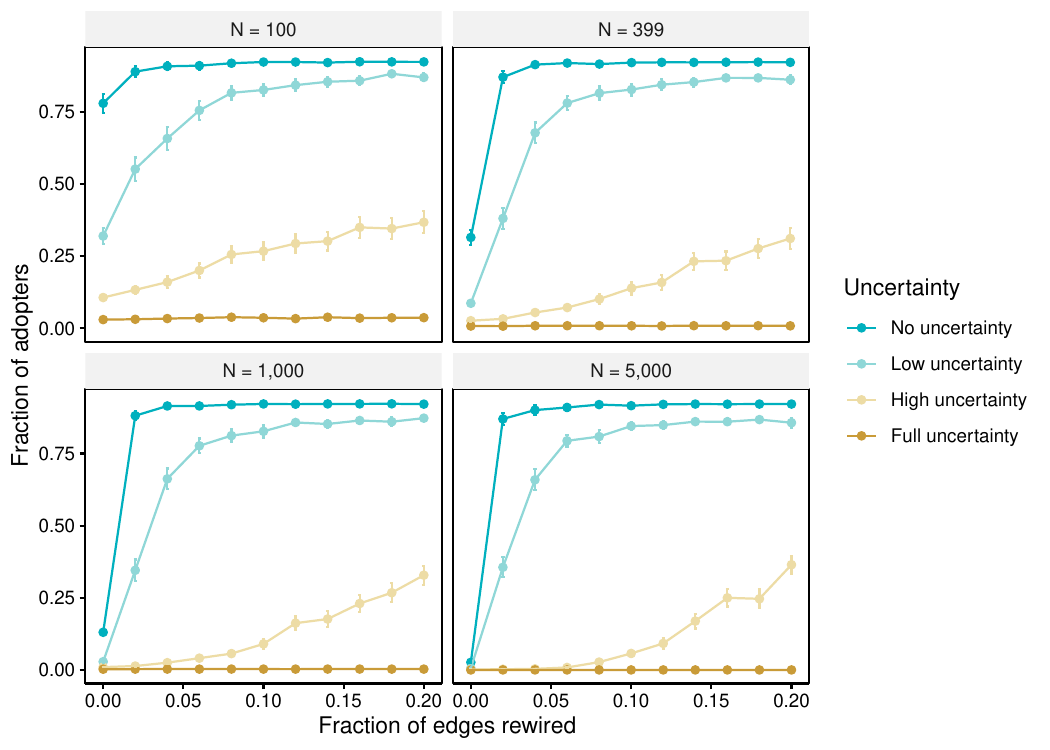}
\caption{\textbf{Robustness check: network size.} Average final fraction of adopters as a function of the fraction of edges rewired, separately for network sizes $N\in\{100,399,1000,5000\}$; the fractional scale places different sizes on a common $x$-axis. High-attractiveness products only; the low-attractiveness case is omitted because diffusion remains negligible at all rewiring levels. For $N>399$, the $399$ participants' thresholds are resampled with replacement to populate the network, a procedure applied uniformly across all sizes. Simulations use ring lattices (degree $k=4$) with degree-preserving rewiring (Sec.~\ref{si:sim-rewiring}); diffusion is seeded from a randomly chosen  connected pair, and thresholds are randomly assigned to nodes each run. Points are means across $R=200$ realizations; error bars are $95\%$ confidence intervals for the mean.}
\label{fig:rob-netsize}
\end{figure}

Long ties raise adoption at every network size (Fig.~\ref{fig:rob-netsize}). The effect is smallest for the $100$-node network, where short paths permit diffusion even without rewiring, and stabilizes for larger networks---confirming that the main-text result is not an artifact of the $399$-node scale.

\clearpage

\subsection{Robustness to seed-set size}\label{si:rob-seed}
\begin{figure}[h]
\centering
\includegraphics[width=.8\linewidth]{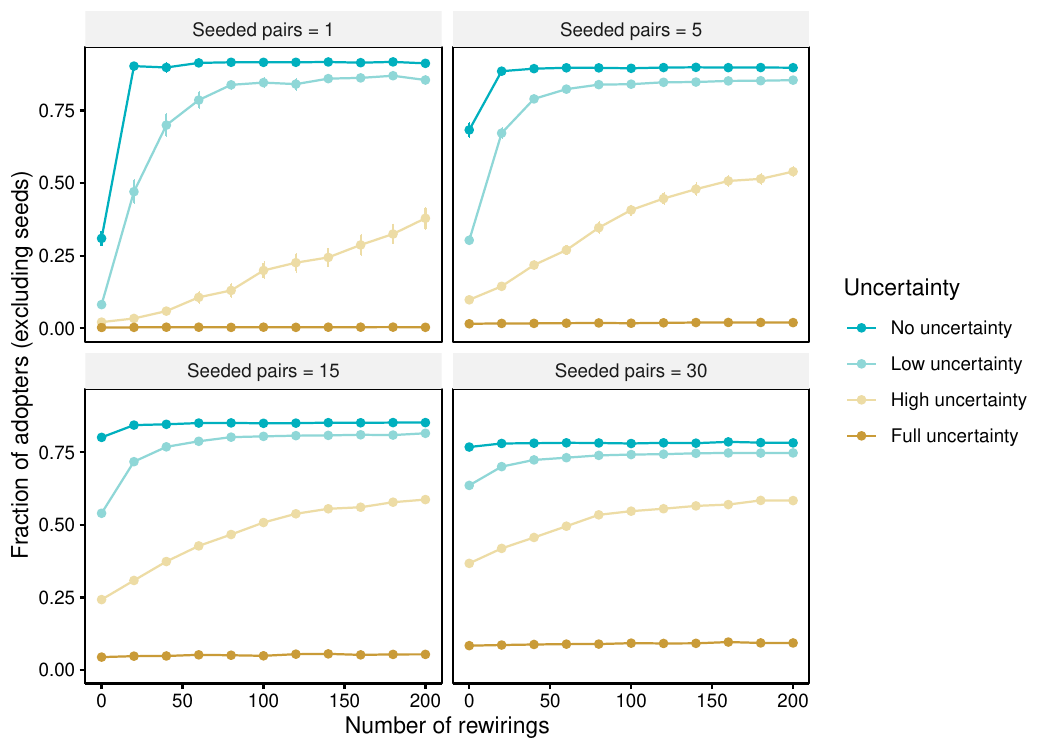}
\caption{\textbf{Robustness check: seed size.} Average final fraction of adopters as a function of the fraction of edges rewired, separately for seed sizes of $1$, $5$, $15$, and $30$ connected pairs. High-attractiveness products only; the low-attractiveness case is omitted because diffusion remains negligible at all rewiring levels. Simulations use ring lattices ($N=399$, degree $k=4$) with degree-preserving rewiring (Sec.~\ref{si:sim-rewiring}); each seed is a randomly chosen connected pair set to adopted, and thresholds are randomly assigned to nodes each run. Points are means across $R=200$ realizations; error bars are $95\%$
confidence intervals for the mean.}
\label{fig:rob-seed}
\end{figure}

Long ties raise adoption at every seed size, but the effect shrinks as the seed set grows (Fig.~\ref{fig:rob-seed}). Larger seed sets spread diffusion further on their own, leaving fewer unexposed low-threshold nodes for long ties to reach---the same headroom logic seen across products, populations, and degree: long ties matter most where baseline diffusion is incomplete.

\clearpage

\section{The no-social-information (baseline) condition}\label{si:baseline}
\begin{figure}[h]
\centering
\includegraphics[width=1\linewidth]{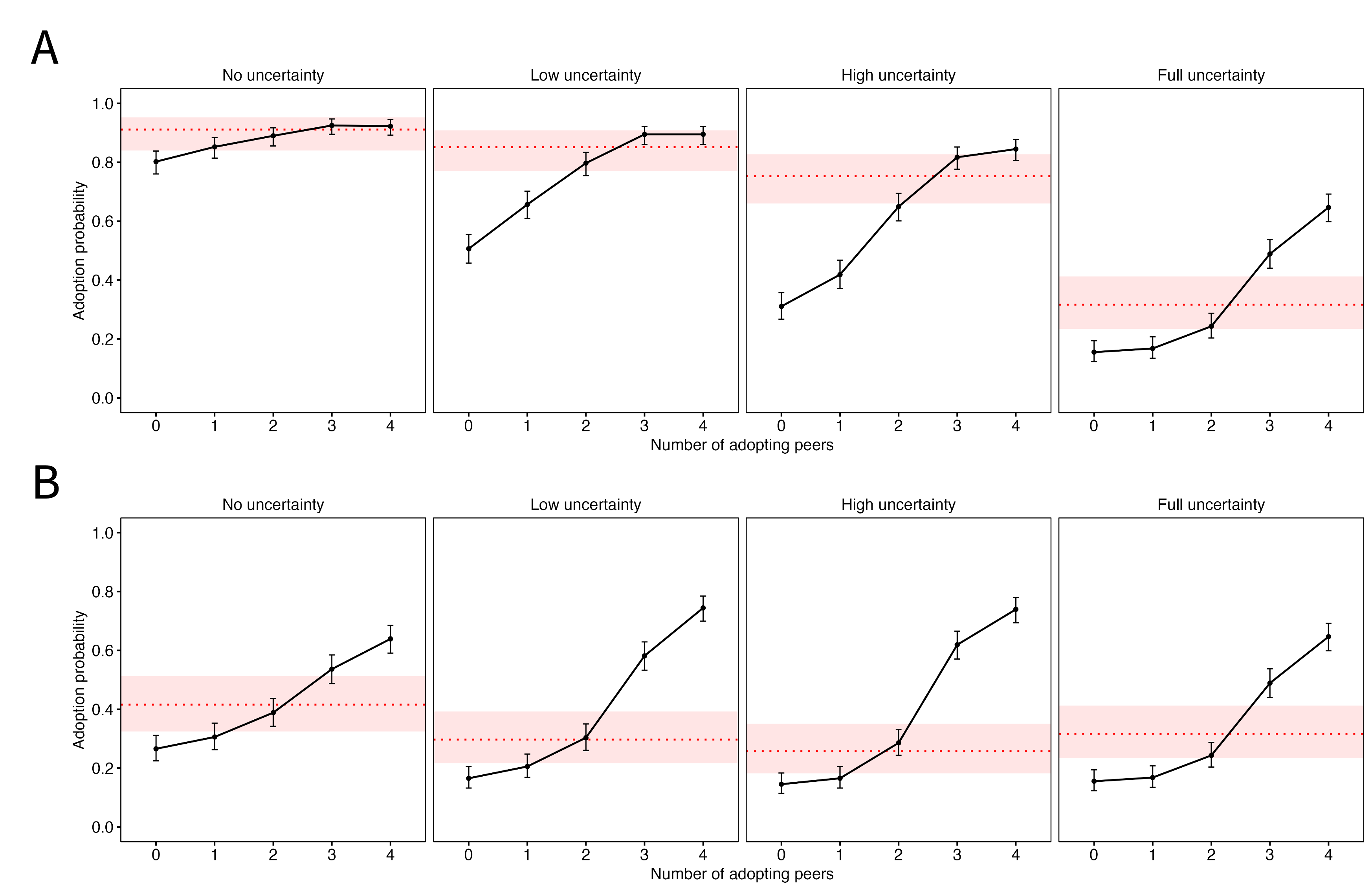}
\caption{\textbf{Adoption probability with and without social information.} (A) High-attractiveness and (B) low-attractiveness products. Within each panel, points show the adoption probability (proportion of participants who adopt, $n=399$) conditional on the number of adopting peers $m\in\{0,\dots,4\}$, for each uncertainty level (no $=0\%$ gray balls, low $=25\%$, high $=50\%$, full $=100\%$). Error bars are $95\%$ confidence intervals for a single proportion. The dashed line marks the baseline adoption probability without social information (independent sample, $n=101$); the shaded band is its $95\%$ confidence interval. Under full uncertainty attractiveness is not manipulated, so the corresponding curves are identical across
panels.}
\label{fig:prob}
\end{figure}

A separate group of participants ($n=101$) completed the same adoption tasks without peer information, providing a baseline estimate of intrinsic adoption propensity. Fig.~\ref{fig:prob} overlays this baseline on the adoption probability in the social-information condition, as a function of the number of adopting peers. 

Two patterns emerge. First, at zero adopting peers, adoption in the social-information condition falls below the baseline in nearly all product configurations: the mere availability of peer information suppresses adoption
relative to deciding without a social frame. This is consistent with the marginal-majority effect documented across social-influence experiments, in which choice probability drops below individual baselines when the majority has not adopted \citep{gelastopoulos2026themarginal}, and with rational social learning, under which observing zero adopters is itself an informative signal.

Second, adoption probability rises with the number of adopting peers and exceeds the baseline, but only for products with high uncertainty or low attractiveness. For high-attractiveness, low-uncertainty products, adoption is near ceiling regardless of peer behavior, leaving little room for social information to operate. This asymmetry aligns with the prediction that social information is most valuable when private signals are weak \citep{toyokawa2019social}. 

\clearpage

\end{document}